# 1. Influence of the time-dependent surfactant adsorption on the lifetime of a drop pressed by buoyancy against a planar interface

Clara Rojas, Máximo García-Sucre and Germán Urbina-Villalba

*Centro de Estudios Interdisciplinarios de la Física, Instituto Venezolano de Investigaciones Científicas (IVIC), Apdo. 20632, Caracas 1020A, Venezuela.*

**Abstract.** Emulsion Stability Simulations (ESS) of deformable droplets are used to study the influence of the time-dependent adsorption on the coalescence time of a 200-μm drop of soybean oil pressed by buoyancy against a planar water/oil interface. The interface is represented by a 5000-μm drop of oil fixed in the space. The movement of the small drop is determined by the interaction forces between the drops, the buoyancy force, and its thermal interaction with the solvent. The interaction forces depend on the surface concentration of surfactant molecules at the oil/water interfaces. Assuming diffusion limited adsorption, the surface excess of the surfactant becomes a function of its apparent diffusion constant, $D_{app}$. Distinct probability distributions of the coalescence time are obtained depending on the magnitude of $D_{app}$. The origin and the significance of these distributions are discussed.

Correspondence/Reprint request: Dr. Germán Urbina Villalba, Centro de Estudios Interdisciplinarios de la Física, Instituto Venezolano de Investigaciones Científicas (IVIC), Apdo. 20632, Caracas 1020A, Venezuela. E-mail: guv@ivic.gob.ve

## Introduction

When a small drop of oil is released in the bulk of the water phase it moves according to Stokes law until it approaches the interface to a sufficiently small distance [1,2]. Then it markedly decelerates due to the increase of the viscous friction in the remaining gap. Commonly, the drop rests at the oil/water interface before merging with its homo-phase, although sometimes it coalesces at the impact.

If the drop retains its spherical shape during the approach to the interface ($R_i$ < 10 μm, where $R_i$ is the radius of the drop $i$), its velocity near the surface can be described by the diffusion tensor of Taylor [1,3]. In this case, an inverse dependence of the coalescence time with the radius of the drop is obtained. Such dependence reproduces the lifetime of small drops of hexadecane stabilized with $β$-casein when they are pressed by buoyancy against a water/hexadecane interface [3,4].

Conversely, large drops significantly deform during their approach to the interface, turning roughly into truncated spheres. The water trapped between the drop and the interface generates an approximately planar oil/water/oil film at the interfacial boundary. The time required for the coalescence of the drop can be computed if the velocity of thinning of the liquid film is known. Reynolds calculated the velocity of thinning of a liquid film between two rigid circular sections [5]:

$$V_{Re} = \frac{2 F h^3}{3\pi \eta r_f^4}, \tag{1}$$

here $F$ is the driving force, $h$ is the closest distance of approach between the sphere and the interface, $η$ is the dynamic viscosity of the surrounding liquid, and $r_f$ is the radius of the film. This radius can be found from the stress balance of the drop at the interface:

$$r_f = \frac{F R_i}{\pi \gamma}, \tag{2}$$

where γ is the interfacial tension.

The time elapsed from the moment in which the drop starts to move slowly until it coalesces with the large homophase ($\tau$) can be calculated if it is assumed that the film drains steadily until coalescence occur:

$$\tau = \int_{h_{crit}}^{h_{ini}} \frac{dh}{v(h)}, \tag{3}$$

here $v(h)$ is the velocity of thinning of the liquid film, $h_{ini}$ is the initial distance of deformation of the drop, and $h_{crit}$ is the critical thickness of rupture: the minimum distance that can be attained before the film breaks and coalescence occurs.

Using Eqs. (1)-(3) an explicit expression of $\tau$ can be obtained:

$$\tau = \frac{3\eta F r_i^2}{4\pi\gamma^2}\left(\frac{1}{h_{crit}^2} - \frac{1}{h_{ini}^2}\right). \tag{4}$$

In the absence of interaction forces, F can be approximated by the buoyancy force:

$$F = \frac{4}{3}\pi r_i^2 \Delta\rho g. \tag{5}$$

Introducing Eq. (5) into Eq. (4) it is found that the average coalescence time of large drops augments as a function of a particle radius.

It is well known that the rate of thinning of a film depends on its exact shape [6]. The exact shape of the film depends on the deformability of the drop and the interface. Owing to the difference in the curvature of the drop and the interface, the drop contacts the interface at a narrow ring-shaped region called the barrier ring [7]. Usually the film is thicker at its center and thinner around the circumference $2\pi r_f$. Hence, Eqs. (1) and (4) are only approximations to the real thinning process.

In general, a distribution of lifetimes -considerably smaller than the ones predicted by Eq. (4)- is observed. This distribution is usually wide, containing drops with extremely short rest times. When the system contains dissolved surfactants, exceptionally long lifetimes are also observed.

The experiments of Gillespie and Rideal showed that in the absence of surfactants the drops of benzene and liquid paraffin (1.5 mm < $R_i$ < 4.5 mm) [8] already show a statistical distribution of coalescence times. Hence, the stochastic nature of the coalescence phenomena should be connected with the process of rupture of the film [2]. Moreover, since the approach of the film-thinning models is deterministic, they cannot account for a distribution of rest times.

According to Gillespie and Rideal, the rupture of film occurs when the amplitude of its capillary waves reaches the width of the film. If $N$ is the number of drops which do not coalesce in a time $t$, and $N_0$ the total number of drops measured in the experiment, the probability of rupture per second $1/N \, (dN/dt)$ of the film is given by:

$$\log(N/N_0) = -K(t-t_0)^{3/2}. \qquad (6)$$

Where:

$$K = f \, C_0 \, A_0 \, (6\gamma / R_i \eta)^{1/3}. \qquad (7)$$

Here: $f$ and $A_0$ correspond to the frequency and the amplitude of the capillary wave. Equation (6) successfully reproduces the data of Gillespie and Rideal [8].

Ghosh and Juvekar [7] suggested an alternative mechanism for the drop rest phenomenon considering the effect of surfactant molecules. According to these authors, the flux of water generated between the incoming drop and the interface displaces the surfactant molecules adsorbed in the outward direction, from the interior of the film towards its borders, promoting a sudden accumulation of surfactant molecules at the barrier ring. This generates a very strong repulsive force which causes the bouncing of the drop at the interface, and its subsequent arrest. The force decays in time due to the back diffusion of surfactant molecules towards the center of the film. The distribution of drop rest times is caused by a fluctuation of the surfactant surface excess at the location where the drop strikes the planar interface. Such fluctuations are caused by: a) the disturbance of the interface resulting from the coalescence of previously added drops, and b) the non-uniform distribution of adsorbate at the barrier ring resulting from the hydrodynamic perturbation of the interface when the drop moves at close proximity. The authors hypothesized that the incoming drop attains equilibrium with the surrounding phase before it rests. Hence the variation of $\Gamma$ only occur at the planar interface.

Ghosh and Juvekar derived an analytical equation for cumulative probability distribution of drop rest times $F(\tau_R)$. This probability depends on a dimensionless coalescence threshold $P_r$, a normalized standard deviation of the surface excess $S_r$, and the effective diffusion coefficient of the surfactant at the oil/water interface, $D_r$. The model should be fitted to the experimental data in order to evaluate $P_r$ and $S_r$. The value of $D_r$ should be estimated due to its coupling with $P_r$. The reproducibility of these parameters is very good except in the absence of surfactant molecules.

Accurate predictions of the *average* lifetime of soybean droplets (10 μm ≤ $R_i$ ≤ 1000 μm) stabilized with Bovine Serum Albumin (BSA) can be obtained using Emulsion Stability Simulations (ESS) [2]. These calculations suppose: (a) the formation of a plane parallel film between the drop and the interface, (b) a constant concentration of surfactant at the oil/water interfaces ($\Gamma$), and (c) a small steric potential between the drop and the planar interface.

## Emulsion Stability Simulations

In ESS [3,9-12] the drops move with an equation of motion similar to the one of Brownian dynamic simulations:

$$\vec{r}_{p,i}(t+\Delta t) = \vec{r}_{p,i}(t) + \frac{D_i \vec{F}_i}{k_B T}\Delta t + \vec{R}, \qquad (8)$$

where $\vec{r}_{p,i}$ is the position of particle $i$, $D_i$ is its diffusion constant, $\vec{F}_i$ is the total force acting on $i$, $k_B$ is the Boltzmann constant, $T$ is the temperature, $\Delta t$ is the time step, and $\vec{R}$ is a random term which represents the Brownian motion of the particle.

The simulations used to reproduce the drop rest phenomenon use a very large drop of oil ($R_j$ = 5000 μm) fixed in the space to simulate the oil phase of the planar interface. A small drop of oil was released 35 μm below the interface and its coalescence time was computed. Average coalescence times result from 300 random walks of the small drop in its path to the interface. The standard deviation of these walks measures the scattering of coalescence times.

In ESS it is assumed that the molecules of oil mainly determine the van der Waals interaction between the drops (Table 1). Instead, the repulsive interactions depend on the amount and chemical nature of the surfactant molecules adsorbed to the interface of the drops. Once the surfactant has been allocated, the surface properties of the drops (such as charge, interfacial tension, etc.) can be computed.

Then, the diffusion constant of the drops and interaction forces between drops can be calculated. This allows to motion movement of the drops according to Eq. (8). At every time step, the program checks for the coalescence of drops.

The model of truncated spheres is used to simulate the change of shape of the drops as they approach to each other [12-14]. Following the model of truncated spheres three regions of approach can be defined:

a) Region I: The distance of separation between the centers of mass of the drops, $r_{ij}$, is larger than $r_i + r_j + h_{ini}$, where $h_{ini}$ is the initial distance of deformation of the drops:

$$h_{ini} = \frac{2 r_i^3 \Delta \rho g}{3 \gamma}. \tag{9}$$

In this region the drops behave as spherical particles.

b) Region II: This region covers the range of distances between the beginning of the deformation, $r_f \neq 0$:

$$r_f^2 = r_i^2 - \left[ \frac{(r_{ij} - h_{ini})^2 - (r_j^2 - r_i^2)}{2(r_{ij} - h_{ini})} \right]^2, \tag{10}$$

and the attainment of the maximum film radius, $r_f = r_{fmax}$:

$$r_{fmax} = f_r r_i^2 \sqrt{\frac{2 \Delta \rho g}{3 \gamma}}, \tag{11}$$

here $f_r$ = 7/20. As soon as the drops enter region II, they change their spherical shape to a truncated spheroid.

**Table 1**. Interaction potentials used in the simulations. In these equations, $r_i$ is the radius of the small droplet, and $r_j$ is the radius of the large drop simulating the homophase. $A_H$ is the Hamaker constant, $h$ the distance of closest approach between the surfaces, $x = h/2r_i$, $y = r_i/r_j$, $l = h + r_i + \sqrt{r_i^2 - r_f^2}$, $L = r_j + \sqrt{r_j^2 - r_f^2}$, $d = r_j + \sqrt{h^2 + 4r_f^2}$, $r_f$ the radius of the film, $\kappa^2 = \left(8\pi e^2 z^2 / \varepsilon k_B T\right) C_{el}$, where $z$ is the valence, $\varepsilon$ is the dielectric permittivity of the medium, $C_{el}$ is the electrolyte concentration, $e$ is the electron charge, $k_B T$ the thermal energy, and $\psi_{si}$ and $\psi_{sj}$ are the surface potentials of the small and large drops, respectively. For the extensional (dilational) and bending potentials, $\gamma_0$ is the interfacial tension, $r_a = 2r_i r_j / (r_i + r_j)$, and $B_0 = 1.6 \times 10^{-12}$ N [15,16].

| | | |
|---|---|---|
| van der Waals potential (spheres) | $V_{vdW} = -\dfrac{A_H}{12}\left[\dfrac{y}{x^2+xy+x} + \dfrac{y}{x^2+xy+x+y} + 2\ln\left(\dfrac{x^2+xy+x}{x^2+xy+x+y}\right)\right]$ | [17] |
| Electrostatic potential (spheres) | $V_{elect} = \dfrac{64\pi}{\kappa} C_{el} K_B T \tanh\left(\dfrac{e\Psi_{si}}{4k_B T}\right)\tanh\left(\dfrac{e\Psi_{sj}}{4k_B T}\right) e^{-kh}\left[\dfrac{2r_i r_j}{\kappa(r_i+r_j)}\right]$ | [14] |
| van der Waals potential (truncated spheres) | $V_{vdW} = -\dfrac{A_H}{12}\left\{\dfrac{2r_j(l-h)}{l(L+h)} + \dfrac{2r_j(l-h)}{h(l+L)} + 2\ln\left[\dfrac{h(l+L)}{l(h+L)}\right] + \dfrac{r_f^2}{h^2} - \dfrac{l-h}{L}\dfrac{2r_i^2}{hl} - \dfrac{l-r_i-(L-r_j)}{2l-2r_i-h}\dfrac{2r_j^2}{hl} \right.$ $-\dfrac{2(L-r_j)-h}{2l-2r_i}\dfrac{d-h}{2h} + \dfrac{2r_j L^2(l-h)}{hl(l+L)(L+h)} - \dfrac{2r_j}{h(2l-2r_i-h)}\dfrac{l^2+r_f^2}{(l+L)(l+L-2r_j)} +$ $+\dfrac{2r_j^2 d}{(2l-2r_i-h)\left[(h+L)(h+L-2r_j)-(l-h)(l-2r_i-h)\right]} -$ $\left. -\dfrac{4r_j^3(l-h)}{(l+L)(l+L-2r_j)\left[(h+L)(h+L-2r_j)-(l-h)(l-2r_i-h)\right]}\right\}$ | [14] |
| Electrostatic potential (truncated spheres) | $V_{elect} = \dfrac{64\pi}{\kappa} C_{el} K_B T \tanh\left(\dfrac{e\Psi_{si}}{4k_B T}\right)\tanh\left(\dfrac{e\Psi_{sj}}{4k_B T}\right) e^{-kh}\left[r_f^2 + \dfrac{2r_i r_j}{\kappa(r_i+r_j)}\right]$ | [14] |
| Dilational potential (truncated spheres) | $V_{dil} = \dfrac{\pi\gamma r_f^4}{2r_a^2}$ | [14] |
| Bending potential (truncated spheres) | $V_{bend} = -\dfrac{2\pi B_0 r_f^2}{r_a}, \quad (r_f/r_a)^2 \ll 1$ | [12] |

c)  Region III: The maximum film radius has been attained, $r_f = r_{fmax}$, and the intervening liquid between the drops drains until it reaches a critical distance of rupture: $h_{crit}$ [18-21]:

$$h_{crit} = \left(\frac{A_H A_{crit}}{128\gamma}\right)^{1/4}, \quad (12)$$

where $A_{crit} = r_f/10$ and $A_H$ is the Hamaker constant.

The potential of interaction and the diffusion constant of the drops correspond to the ones of spherical particles within region I. At $h = h_{ini}$, the code calculates the dimensions of truncated spheres which are compatible with the actual distance of separation between the centers of mass of the spherical drops. The tensor of Danov *et al.* [13] is used to move the drops:

$$D_{Da} = \frac{4h}{r_i}\left(1 + \frac{r_f^2}{r_i h} + \frac{\varepsilon_S r_f^4}{r_i^2 h^2}\right)^{-1} D_0. \quad (13)$$

The parameter $\varepsilon_S$ was fixed to 1.0 in order to simulate the behavior of tangentially immobile interfaces.

The occurrence of capillary waves is simulated in the program using:

$$\lambda_{TOTAL} = (\lambda_i + \lambda_j)\left[\exp\left(\frac{\tau_{ij}}{\tau_{Vrij}}\right) - 1\right], \quad (14)$$

where $\tau_{ij}$ is the lifetime of the film, and $\lambda_k$ is equal to:

$$\lambda_k = Ran(t) h_{crit}. \quad (15)$$

Here $Ran(t)$ is a random variable between -1.0 and 1.0. $\tau_{Vrij}$ is given by the analytical expression of Vrij and Overbeek for the fastest increase of surface oscillations in a film of width $h_0$ [22,23]:

$$\tau_{Vrij} = 96\pi^2 \gamma \eta h_0^5 A_H^{-2}. \quad (16)$$

The simulations of Ref. [2] confirmed that neither the thinning of a plane-parallel film or the occurrence of capillary waves gives rise to a large dispersion of coalescence times if the surface excess of the surfactant is assumed to be constant during the approach of the drop to the interface [2].

The mechanism of capillary waves only favors a high dispersion of rest times if it is assumed that the adsorption of surfactant molecules to the oil/water interfaces is time-dependent ($\Gamma = \Gamma(t)$), and a high steric potential is generated as a result of the surfactant adsorption.

The implementation of time-dependent adsorption in the ESS program assumes that the surfactant (BSA protein in this case) adsorbs to the oil/water interface following a diffusion limited adsorption mechanism [24-27]. Hence:

$$\Gamma(t) = 2\left(D_{app}/\pi\right) C_p t^{1/2}. \quad (17)$$

Where $C_p$ is the protein concentration. As the surface excess of the protein increases, the value of the interfacial tension decreases from 50 mN/m to 15 mN/m. Hence, $h_{ini}$, $r_f$, and the interaction potentials change as a function of time. The time required for a complete coverage of the oil/water interface ($t_c$) is given by:

$$t_c = \left(\Gamma_{max}/A_c C_p\right)^2. \quad (18)$$

In the present case: $\Gamma_{max} = 9.08 \times 10^{15}$ proteins/m$^2$, and $A_c = 2 \, (D_{app}/\pi)^{1/2}$.

On the one hand, the occurrence of a time-dependent adsorption is likely to happen due to the characteristics of the experimental set up. A drop of oil is generated in the bulk of an aqueous surfactant solution. As soon as the drop is formed it approaches the interface due to the buoyancy force. At the same time the surfactant dissolved starts to adsorb to the interface of the drop. Hence, it is very likely that the surfactant population adsorbed varies from one drop to another. Different values of $\Gamma(t)$ promote dissimilar random paths (the repulsive potential depends on $\Gamma(t)$). The mechanism of rupture by capillary waves depends on the width of the film and on the interfacial tension of its oil/water boundaries. Consequently, different values of $\Gamma(t)$ produce a distribution of coalescence times.

On the other hand, the occurrence of a high steric barrier between the drop and the interface is more difficult to justify. A high repulsive potential is necessary in order to slow down the movement of the incoming drop so that the growth of the capillary waves is possible. While the potentials of Table 1 are easily parameterized, the steric potential produced by a large protein like BSA is difficult to characterize. Hence, the parameters of this potential (potential I in Table 2) were adjusted in order to reproduce the behavior of small drops of the same system reported by Basheva et al. [1]. The magnitude of the resulting potential is small in comparison to the van der Waals attraction. As a consequence, the total potential of interaction between the drop and the interface is only slightly repulsive (Figure 1). A dispersion of coalescence times is not obtained unless a "harder" steric potential (potential II in Table 2) is employed in the simulations.

An additional source of uncertainty regarding the repulsive potential is related to the magnitude of the constant $B_0$ of the bending potential. Constant $B_0$ contains the spontaneous curvature of the surfactant at the interface, a parameter which is difficult to estimate. The value $B_0$ employed in the simulations is the one previously used for reproducing the data of Hofmann and Stein [28] regarding the flocculation of decane-in-water emulsions stabilized with bis-ethyl-hexyl sulfosuccinate (AOT) [29]. This value is one order of magnitude lower than the theoretical estimation [30]. A higher value of $B_0$ would favor the appearance of a large repulsive barrier in the total potential of interaction between the drop and the interface. This barrier is not present in the case of spherical drops, because the bending potential only acts during the deformation of the drops. Consequently, it was not taken into account in the parameterization of the steric potential (I) used in the simulation of small spherical drops [2].

## Discussion and Results

According to ESS the mechanism of time-dependent adsorption does not produce a dispersion of lifetimes when the coalescence occurs via the thinning of the film [2]. An extremely low value of the diffusion constant of the surfactant, $D_{app}$ ($\sim 10^{-12}$ m$^2$/s) causes a smooth approach of the drop to the interface until coalescence occurs. This happens because the surfactant molecule is too slow to adsorb significantly to the surface of the drop before it reaches the interfacial boundary. Hence, an insufficient repulsive barrier is formed. Conversely, an exceptionally high value of $D_{app}$ ($\sim 10^{-7}$ m$^2$/s) generates a repulsive barrier (rb) very quickly, at a long distance of approach. Consequently, the drop stops at the outermost boundary of the repulsive potential ($h_{rb} \sim 30$ nm, Fig. 1), and does not coalesce. Intermediate values of $D_{app}$ ($\sim 2.9 \times 10^{-9}$ m$^2$/s) favor a closer approach of the drop to the interface. The drop penetrates substantially into the repulsive barrier ($h \sim 12$ nm $<< h_{rb}$), before it is pushed outwards abruptly due to the large magnitude of the repulsive steric force at a short distance of separation.

**Table 2.** Steric Potentials used in the simulations. $V_w$ is the molar volume of the solvent, $\chi$ is the Flory-Huggins solvency parameter, $\bar{\phi}_i$ and $\bar{\phi}_j$ are the average volume fraction of the protein around the sphere and the interface, $\Gamma$ is the number of protein molecules per unit area $(9.08 \times 10^9 \text{ moleculas/m}^2)$, $\rho_p$ is the density of the protein, $\delta$ is the width of the protein layer, $v_a$, $v_b$, and $v_c$ are volumes whose explicit expressions are given in Ref. [31], $h_g = h/(2L_g)$, $L_g = N_{seg}\left(\Gamma l_{seg}^5\right)^{1/3}$, $N_{seg}$ is the number of segments of the protein (580 amino acids [31]), $l_{seg}$ is the segment length $(\approx 3.0 \times 10^{-10}\text{m})$. The value of $l_{seg}$ was approximated by $V_a^{1/3}$, where $V_a$ is the typical volume of an amino acid residue ($V_a = 57 - 186$ Å$^3$ [32]).

| | | |
|---|---|---|
| Steric Potential I | $V_{stI} = \dfrac{4k_BT}{3V_1}\bar{\phi}_i\bar{\phi}_j\left(\dfrac{1}{2}-\chi\right)\left(\delta-\dfrac{h}{2}\right)\left[\dfrac{3(r_i+r_j)}{2}+2\delta+\dfrac{h}{2}-\dfrac{3(r_j-r_i)^2}{2(h+r_i+r_j)}\right],\quad \delta<h<2\delta$ | [31,33] |
| | $V_{stI} = \dfrac{k_BT}{V_1}\left(\dfrac{1}{2}-\chi\right)\left[(\bar{\phi}_j)^2\left(\dfrac{v_a^2}{v_c}-v_a\right)+(\bar{\phi}_i)^2\left(\dfrac{v_b^2}{v_c}-v_b\right)+2\bar{\phi}_i\bar{\phi}_j\left(\dfrac{v_av_b}{v_c}\right)\right],\quad 0<h<\delta$ | [31] |
| Steric Potential II | $V_{stII} = \pi r_f^2 f(h) + 4\pi r_i k_B T \Gamma^{3/2} L_g^2 \left(1.37 h_g - 0.2 h_g^{11/4} + 3.20 h_g^{-1/4} - 4.36\right)$ $f(h) = 2k_BT\Gamma^{3/2}L_g\left(\dfrac{4}{5}h_g^{-5/4}+\dfrac{4}{7}h_g^{7/4}-1.37\right)$ | [34,35] |

As can be inferred from the previous paragraph, the modification of the standard film thinning mechanism due to the time-dependent surfactant adsorption does not generate a significant scattering of lifetimes. Either the drop coalesces very fast or does not coalesce. Coalescence only occurs at a low surfactant surface excess.

When the mechanism of capillary waves is added to the ones of film-thinning and time-dependent adsorption, it guarantees that the coalescence of the drop will occur despite the magnitude of the repulsive potential between the drop and the interface. According to our simulations it is the combination of these three processes which reproduces the experimental behavior.

In order to illustrate the effect of the capillary waves, four intermediate values of $D_{app}$ were chosen (Fig. 2). A (small) value of $2 \times 10^{-9}$ m$^2$/s favors the rapid thinning of the film, but in this case the film breaks at very short distances (6-8 nm) due to capillary waves. The process can be described in detail as follows:

1) The lifetime of the film $\tau_{ij}$ increases from zero as soon as the drop deforms and the film forms ($h = h_{ini}$). 2) Since the values of $h_0$ (Eq. (16)) correspond to the thickness of the film, then $h_0 = h$ in the regions II and III of the deformation process. Hence $\tau_{Vrij}$ decreases as the drop approaches the interface. 3) At very short distances $\tau_{ij}$ reaches values comparable to $\tau_{Vrij}$ ($\tau_{ij} / \tau_{Vrij} \sim 0.4$). At this point the capillary waves increase exponentially and the film ruptures.

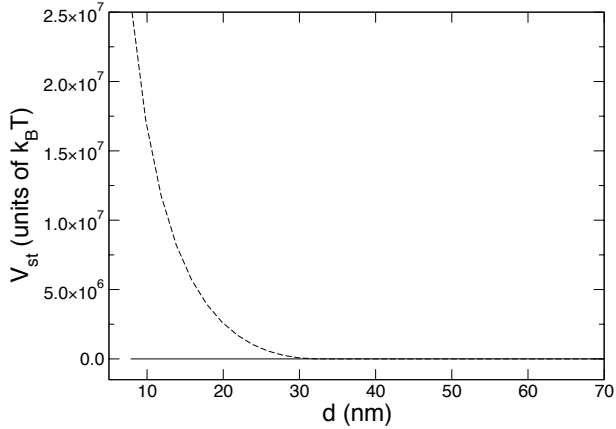

**Figure 1.** Steric potential. Solid line: steric potential I; dashed line: steric potential II with $l_{seg} = 3.00$ Å.

Figure 2(a) illustrates that the cumulative probability of coalescence $P_c$ for $D_{app} = 2 \times 10^{-9}$ m$^2$/s. The typical form of the curves reported by Ghosh and Juvekar for $F(\tau_R)$ [7] is reproduced. The spread of the curve in the abscissa is narrow, and the standard deviation of the simulations is low ($\sigma \sim 0.006$). The corresponding probability of film survival ($1-P_c$) is shown in Fig. 3(a). A logarithmic scale is used to allow a direct comparison with the plots of Gillespie and Rideal [8]. It is clear that the typical form of the curves reported by Gillespie and Rideal [8] is obtained. In summary, having a slow surfactant is equivalent to having no surfactant at all. As Figs. 4(a) and 5(a) demonstrate, h diminishes monotonously in this case until it is equal to zero.

At the other extreme, a high intermediate value of $D_{app} \sim 3.0 \times 10^{-9}$ m$^2$/s generates a sufficiently strong repulsive potential to repel the drop before the film reaches very small widths. Hence, $\tau_{Vrij}$ cannot get small enough during the approach of the drop to be comparable to $\tau_{ij}$. The drop bounces from a distance of approximately 11 nm (Fig. 5(d)), and from that point the film starts to grow. As the drop separates from the interface both $\tau_{ij}$ and $\tau_{Vrij}$ increase. Since $D_{app}$ is fast, the repulsive potential increases with time progressively attaining its maximum strength ($\varGamma = \varGamma_{max}$). The drop is repelled up to $h = h_{rb}$ and stays at its equilibrium distance (Figs. 4(d) and 5(d)) until the film breaks stochastically ($\tau_{ij} / \tau_{Vrij} \sim 0.4$). Again, the plots of Ghosh and Juvekar [7] (Fig. 2(d)) and Gillespie and Rideal (Fig. 3(d)) are reproduced. As in the case of $D_{app} = 2.0 \times 10^{-9}$ m$^2$/s, the standard deviation of the simulations is small but now the film breaks around 17.5 s.

The degree of penetration of the drop into the range of the repulsive potential between the drop and the interface is a function of $D_{app}$. Lower values of the diffusion constant favor weaker repulsive potentials and thinner films. If the minimum separation attained is short enough, the film ruptures at short times while $\tau_{Vrij}$ is decreasing. Otherwise, the drop bounces from the interface. From that point on, $\tau_{Vrij}$ increases asymptotically reaching a constant value at $h = h_{rb}$ and $\varGamma = \varGamma_{max}$ ($\gamma = 15$ mN/m). Eventually $\tau_{ij} \sim \tau_{Vrij}$ and the film breaks. As shown in Figs. 2(b,c) and 3(b,c), the probability distributions do not show intermediate coalescence times between 3 s and 16 s. Either the film ruptures at short times or at large times. The average lifetime increases as $D_{app}$ increases. A large scattering of coalescence times occurs when: $2.86 \times 10^{-9}$ m$^2$/s $\leq D_{app} \leq 2.89 \times 10^{-9}$ m$^2$/s. The maximum standard deviation is obtained for $D_{app} = 2.875 \times 10^{-9}$ m$^2$/s.

According to the experimental data of Ghosh and Juvekar, $F(\tau_R)$ shows a continuous distribution of lifetimes independently of the surfactant concentration used. No gap is observed between short and long lifetimes. The main reason for this discrepancy is the use of Eq. (16) to estimate the fastest increase of surface oscillations, $\tau_{Vrij}$. Equation (16) was deduced assuming the effect of van der Waals forces alone [22,23]. For additional interaction potentials, the expression of Vrij involves a second differential of the potential as function of the film width for short separation distances [23]. This derivative is very difficult to estimate. Moreover, the value of $\tau_{Vrij}$ (Eq. (16)) will now change as the function of the interaction potential besides $h$ and $\gamma$. This would not be a problem except for the fact that the potential is a function of the interfacial properties of the film including $\varGamma(t)$ and $\gamma$. In any event, it is clear that if $\tau_{Vrij}$ changes as a function of the repulsive potential, a variety of conditions for film rupture ($\tau_{ij} \sim \tau_{Vrij}$) can be obtained.

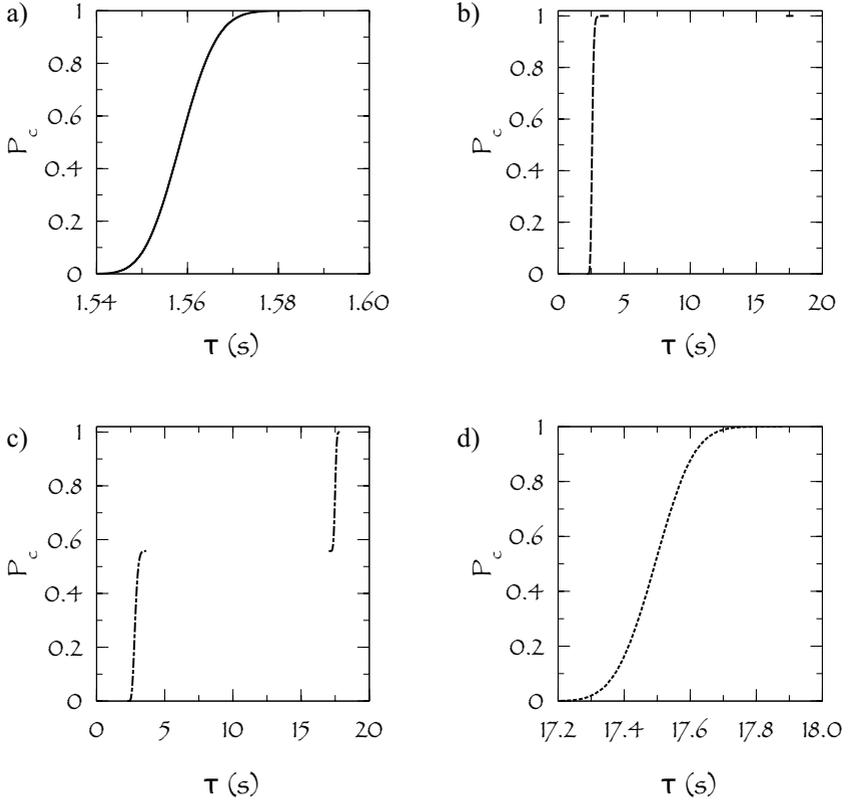

**Figure 2.** Cumulative probability distribution of coalescence times for $R_i = 200$ μm. (a) solid line $D_{app} = 2.00 \times 10^{-9}$ m$^2$/s, (b) dashed line $D_{app} = 2.85 \times 10^{-9}$ m$^2$/s, (c) dotted-dashed line $D_{app} = 2.875 \times 10^{-9}$ m$^2$/s, and (d) dotted line $D_{app} = 3.00 \times 10^{-9}$ m$^2$/s.

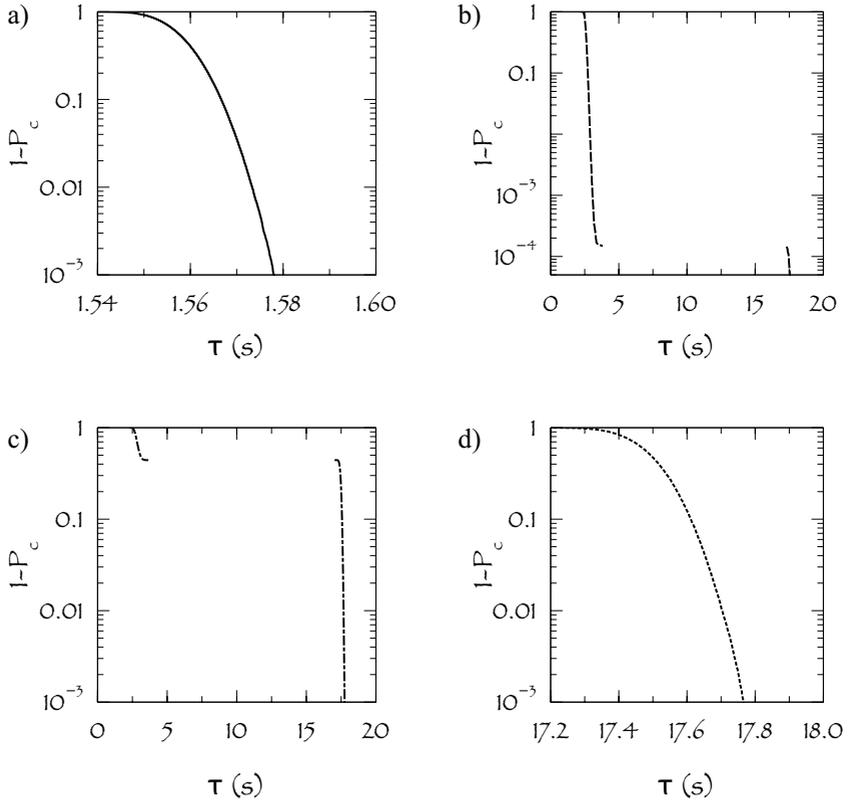

**Figure 3.** Probability distribution of lifetimes (survival times) for a drop of $R_i = 200$ μm. (a) Solid line $D_{app} = 2.00 \times 10^{-9}$ m²/s, (b) dashed line $D_{app} = 2.85 \times 10^{-9}$ m²/s, (c) dotted-dashed line $D_{app} = 2.875 \times 10^{-9}$ m²/s, and (d) dotted line $D_{app} = 3.00 \times 10^{-9}$ m²/s.

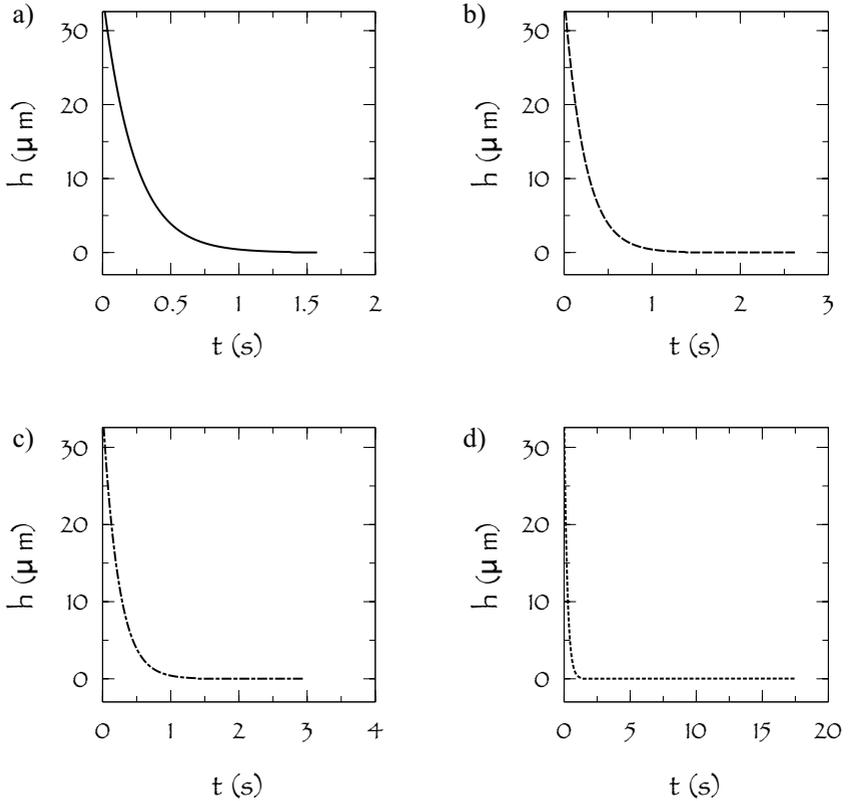

**Figure 4.** Separation distance between the drop and the interface as a function of time for a drop of $R_i$ = 200 μm.  (a) Solid line $D_{app} = 2.00 \times 10^{-9}\, m^2/s$, (b) dashed line $D_{app} = 2.85 \times 10^{-9}\, m^2/s$, (c) dotted-dashed line $D_{app} = 2.875 \times 10^{-9}\, m^2/s$, (d) dotted line $D_{app} = 3.00 \times 10^{-9}\, m^2/s$. The lines end where coalescence occurs.

Time-dependent surfactant adsorption 17

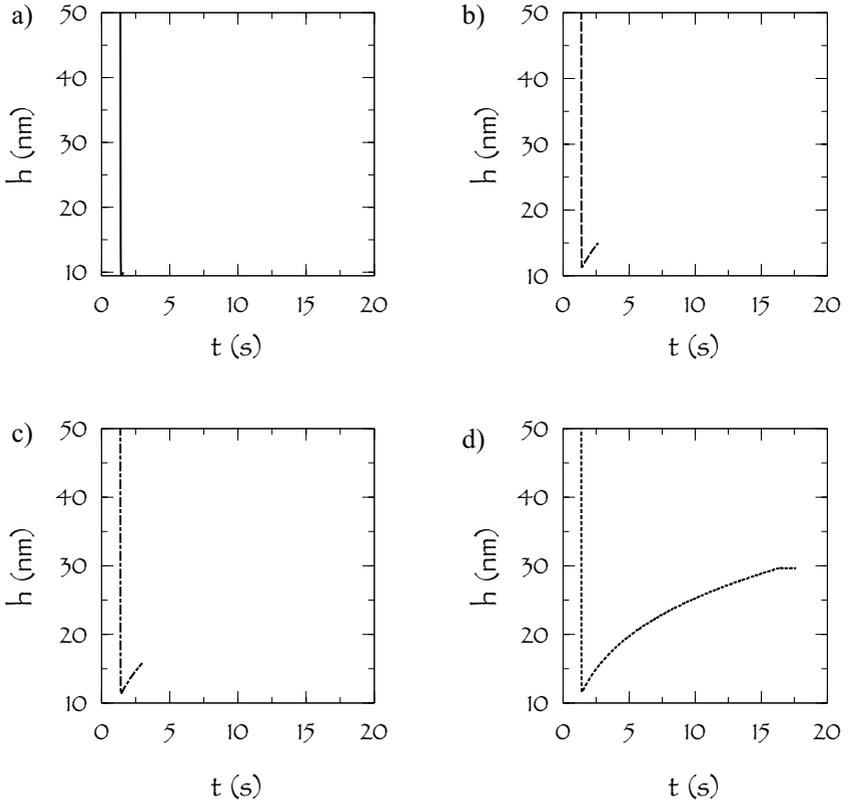

**Figure 5.** Detailed behavior of the drop (Fig. 4) at short distances of separation. (a) Solid line $D_{app} = 2.00 \times 10^{-9}$ m$^2$/s, (b) dashed line $D_{app} = 2.85 \times 10^{-9}$ m$^2$/s, (c) dotted-dashed line $D_{app} = 2.875 \times 10^{-9}$ m$^2$/s, (d) dotted line $D_{app} = 3.00 \times 10^{-9}$ m$^2$/s.

## Conclusion

Emulsion Stability Simulations can be very useful tool for the study of the drop rest phenomenon. According to the simulations the scattering of lifetimes experimentally observed, is due to a combination of the mechanisms of film thinning, time-dependent adsorption and capillary waves. These processes are coupled dynamically due to the dependence of their variables on the interfacial properties of the film formed between the drop and the interface. This properties depend on the surfactant surface excess whose effect can be simulated assuming a diffusion-controlled adsorption with an apparent diffusion constant for surfactant adsorption.

## References

1. Basheva E. S., Gurkov T. D., Ivanov I. B., Bantchev G. B., Campbell B., Borwankar R. P. 1999, *Langmuir,* 15, 6764.

2. Rojas, C., García-Sucre M., Urbina-Villaba G. 2010, *Phys. Rev. E*, 82, 056317.

3. Rojas, C., Urbina-Villaba G., García-Sucre M. 2010, *Phys. Rev. E*, 81, 016302.

4. Dickinson E., Murray B. S., Stainsby G. 1988, *J. Chem. Soc., Faraday Trans. 1*, 84, 871.

5. Reynolds O. 1895, *Phil. Trans. R. Soc. London A*, 186, 123.

6. Mackay G. D. M., Mason S. G. 1963, *Can. J. Chem. Eng.*, 41, 203.

7. Ghosh P., Juvekar V. A. 2002, *Trans IChemE*, 80, 715.

8. Gillespie T., Rideal E. K. 1956, *Trans. Faraday Soc.*, 52, 173.

9. Urbina-Villalba G. , García-Sucre M. 2000, *Langmuir,* 16, 7975.

10. Urbina-Villalba G., Toro-Mendoza J., Lozsán A., García-Sucre M. 2004, *Emulsions: Structure Stability and Interactions,* Elsevier, New York, pp. 677–719.

11. Urbina-Villalba G. 2009, *Int. J. Mol. Sci.*, 10, 761.

12. Toro-Mendoza J., Lozsán A., García-Sucre M., Castellanos Aly J., Urbina-Villalba G. 2010, *Phys. Rev. E,* 81, 011405.

12. Ivanov I. B., Danov K. D., Kralchevsky P. A. 1999, *Colloids Surf. A*, 152, 161.

13. Danov K. D., Denkov N. D., Petsev D. N., Ivanov I. B., Borwankar R. 1993, *Langmuir,* 9, 1731.

14. Danov K. D., Petsev D. N., Denkov N. D., Borwankar R. 1993, *J. Chem. Phys.*, 99, 7179.

15. Kralchevsky P. A., Gurkov T. D., Ivanov I. B. 1991, *Colloid Surf.*, 56, 149.

16. Kralchevsky P. A., Gurkov T. D., 1991, *Colloid Surf.*, 56, 101.

17. Hamaker H. C. 1937, *Physica (Amsterdam),* 4, 1058.

18. Sheludko A. 1967, *Adv. Colloid Interface Sci.*, 1, 391.

19. Ivanov I. B., Radoev B., Manev E., Scheludko A. 1970, *Trans. Faraday Soc.*, 66, 1262.

20. Manev E. D., Nguyen A. V. 2005, *Adv. Colloid Interface Sci.*, 114-115, 133.

21. Manev E. D., Angarska J. K. 2005, *Colloids Surf. A*, 263, 250.

22. Vrij A., Overbeek, J. Th. G. 1968, J. *Ame. Soc.,* 90, 3074.

23. Vrij, A. 1966, *Discuss. Faraday Soc.*, 42, 23.

24. Ward A. F. H., Tordai L. J. 1946*, J. Chem. Phys.*,14, 453.

25. Rosen M. J., Hua X. Y. 1990, *J. Colloid Interface Sci.*, 139, 397.

26. Hua X. Y., Rosen M. J. 1988, *J. Colloid Interface Sci.*, 124, 652.

27. Hua X. Y., Rosen M. J.1991, *J. Colloid Interface Sci.*, 141, 180.

28. Hofman J. A. M. H., Stein H. N. 1991, *J. Colloid Interface Sci.*, 147, 508.

29. Osorio P., Urbina-Villalba G. 2010, *Journal of Surfactants and Detergents*, 14, 281.

30. Kralchevsky P. A., Gurkov T. D., Nagayama K. 1996, *J. Colloid Interface Sci.,* 180, 619.

31. Loszán A., García-Sucre M., Urbina-Villalba G. 2006, *J. Colloid Interface Sci.*, 299, 366.

32. Creighton T. E. 1984, *Proteins* (W. H. Freeman and Company, San Francisco).

33. Loszán A., García-Sucre M., Urbina-Villalba G. 2005, *Phys. Rev. E,* 72, 061405.

34. Alexander S. J. 1977, *J. Phys. (France),* 38, 983.

35. de Gennes P. G. 1987, *Adv. Colloid Interface Sci.*, 27, 189.